\documentclass[twocolumn,epjc3]{svjour3}
\usepackage[utf8]{inputenc}
\usepackage{amsmath,bm}

\usepackage{graphicx,srcltx,amsmath,color,hyperref}

\def\ep{\varepsilon}
\def\eps{\epsilon} 
\def\pa{\partial}

\newcommand{\beq}{\begin{equation}}
\newcommand{\eeq}{\end{equation}}
\newcommand{\beqa}{\begin{eqnarray}}
\newcommand{\eeqa}{\end{eqnarray}}
\newcommand {\apgt} {\ {\raise-.5ex\hbox{$\buildrel>\over\sim$}}\ }

\newcommand{\fract}[2]{{\textstyle\frac{#1}{#2}}}
\newcommand{\fracd}[2]{{\displaystyle\frac{#1}{#2}}}
\newcommand{\der}[2]{\frac{\partial #1}{\partial #2}}

\newcommand{\vb}[1]{\mbox{\boldmath $#1$}} %
\definecolor{myblue}{rgb}{0,0,.7}
\newcommand{\corr}[1]{{ #1}}

\newcommand{\cP}{{\cal P}}
\newcommand{\dP}{{\partial \cal P}}
\newcommand{\cN}{{\cal N}}
\newcommand{\cC}{{\cal C}}

\newcommand{\vbx}{\vb{x}}
\newcommand{\dd}[1]{\delta^{#1}}
\def\oht{{\textstyle \frac{1}{2}}}
 
\newcommand{\ii}{\mathrm{i}}
\newcommand{\sumprime}{\sideset{}{'}{\sum}}
\newcommand{\Ac}{{\cal A}}
\newcommand{\drho}{{\Delta\rho}}

\graphicspath{{figs/}}

\begin{document}

\journalname{Eur. Phys. J. A}

\title{Shape stability of pasta phases: Lasagna case}

\author{Sebastian Kubis\thanksref{a,addr} \and W\l{}odzimierz~W\'ojcik\thanksref{addr}}

\institute{Institute of Physics, Cracow University of Technology,  Podchor\c{a}\.zych 1, 30-084 Krak\'ow, Poland
          \label{addr}}

\thankstext{a}{email: skubis@pk.edu.pl}

\maketitle

\begin{abstract}

The stability of periodically placed slabs occurring in neutron stars (lasagna  phase)  is 
examined by exact geometrical methods for the first time. It appears that the slabs are  stable against any shape perturbation modes for the whole range of volume fraction occupied by the slab. The calculations are done in the framework of the  liquid drop  model and obtained results are universal -  they do not depend on model parameters  like surface tension or charge density. The results shows that the transition to other pasta shapes requires crossing the finite energy barrier.\\

\end{abstract}

\section{Introduction}

The very long  structures appearing in neutron star matter, called pasta phases, 
are a commonly accepted { phenomenon}. Following the seminal work \cite{ravenhall83}, 
the pasta phases have been studied in many different manners.
Many approaches are based on the compressible liquid drop model (CLDM) 
where the system is described by two homogeneous phases separated by a sharp boundary with non-zero 
surface tension \cite{Baym:1971ax,hashimoto,Oyamatsu:1993zz,Williams:1985prf}. The competition between the Coulomb and surface energy leads to different shapes which are usually
described in the Wigner-Seitz approximation, where the geometry of the phase is  imposed at the beginning.
The spherical and cylindrical cells are not the correct {\em unit cells} 
as they are not able to fill the whole space by periodic placement. Structures obeying periodicity, in the form of gyroid and diamond-like shapes, were introduced in \cite{Nakazato:2009ed,Nakazato:2010nf}.  
However, they must be treated only as
an approximation of a true shape, because they do not satisfy the necessary condition for the cell energy extremum. It was shown in \cite{Kubis:2016fmw} that the condition relates the mean curvature $H(\vb{x})$ of the cluster surface
and electrostatic potential $\Phi(\vb{x})$
\beq
2\sigma H(\vb{x})= {\cal C} + \Delta\rho\;\Phi(\vb{x})~,
\label{H-eqn}
\eeq 
where $\sigma$ and $\Delta\rho$ are surface tension and charge difference between the cluster and its  
surroundings.The constant  $\cal C$
depends on pressure difference between neutron and proton phases and mean value of potential 
${\cal C }= P^\cN - P^\cP + \Delta\rho\; \langle \Phi\rangle_\cP $.
It  means the surface curvature  depends on the potential distribution and is not constant 
in general. In fact, the phases considered in \cite{Nakazato:2009ed}
represent surfaces with $H\!=\!0$ and as such they cannot be the true solution of
Eq.~(\ref{H-eqn}).
In works \cite{Nakazato:2009ed,Nakazato:2010nf} it was also shown that they have larger energy than the flat slabs.
Nevertheless, the energy difference is not large, so it seems that such triple-periodic structures are likely to occur.

Recent analysis of pasta phases, including the periodicity, based on  quantum or classical Molecular Dynamics
 \cite{Watanabe:2004tr,Schneider:2013dwa,Schneider:2014lia,Horowitz:2015gda,Alcain:2014fma} 
and Time-Dependent Hartree-Fock \cite{Schuetrumpf:2012cj,Schuetrumpf:2014aea} or Hartree-Fock
with twist-averaged boundary conditions \cite{Schuetrumpf:2015nza}
have shown the existence of triply-periodic ones in the form of gyroid, diamond, twisted spaghetti and  waffles. 
Topological analysis of such structures was presented in  \cite{Kycia:2017ibr}.

The general motivation of our work is to confirm the existence of such non-trivial periodic structures 
also in the framework of CLDM. 
Therefore, we propose to examine the possibility  of transition from lasagna phase to other 
shapes by its shape deformation. The inspection of various perturbation modes could indicate into what kind of 
structures the lasagna is going to transform. Such consideration corresponds to the stability 
analysis of the lasagna phase. The answer to the stability question is not obvious.  Though the contribution from 
surface energy for flat slab is always positive, the Coulomb energy is not and could destabilize the lasagna.

\corr{
Another type of considerations, connected to the stability issue, were presented in the works 
\cite{DiGallo:2011cr,Urban:2015dna,Kobyakov:2018ftm,Durel:2018cxs} where the hydrodynamic approach was used to 
describe the 
 modes with the density perturbation in pasta phases. The obtained mode frequencies  were always real which means that  the  lasagna phase is stable for this kind of collective modes. Our approach concerns a different kind of excitations - it  considers only the shape-changing modes while the nucleon density  is kept constant. Such analysis makes the overall  stability discussion more complete.}

In the work \cite{Kubis:2016fmw} an overall discussion of the rigorous treatment of periodically placed proton clusters of any shape was presented. 
The stable cluster surface should satisfy not only the Eq.~(\ref{H-eqn}), which represents the necessary condition for the  extremum, but also the condition for the minimum coming from the inspection of second variation of the energy functional. 
The second order analysis, in a limited sense, was already carried  for cylinders and balls  in the  \cite{Iida:2001xy}, where a particular deformation mode
was examined in the isolated Wigner-Seitz cell. 
 Periodically placed cylinders and balls were considered in \cite{Williams:1985prf} and later in  
 \cite{Pethick:1998qv}, 
but we need to be aware that these structures
do not represent the proper minimum determined by the Eq.~(\ref{H-eqn}), their shape was assumed apriori.
Up to now, the only known true periodic solution of Eq.~(\ref{H-eqn}) is lasagna (the boundaries of slabs
coincide with the equipotential surfaces) and thus  the stability analysis of these structure may be carried out

Partial stability analysis of lasagna phase was done by Pethick and Potekhin in the context of elastic properties of the phase in the work \cite{Pethick:1998qv}.  In order to determine the elasticity coefficient, they considered 
one particular deformation mode and moreover the expression for the energy was valid only in the limit of wavelength going to infinity. Such analysis corresponds to the stability consideration, however, being limited  to the only one type of deformation. Here, we present general, unconstrained stability analysis for any kind of deformation with finite wavelength, which, to the authors' knowledge, has never been carried out.

\section{Energy variation for single slab}

\begin{figure}[t!]
\centering
\includegraphics[width=.85\columnwidth]{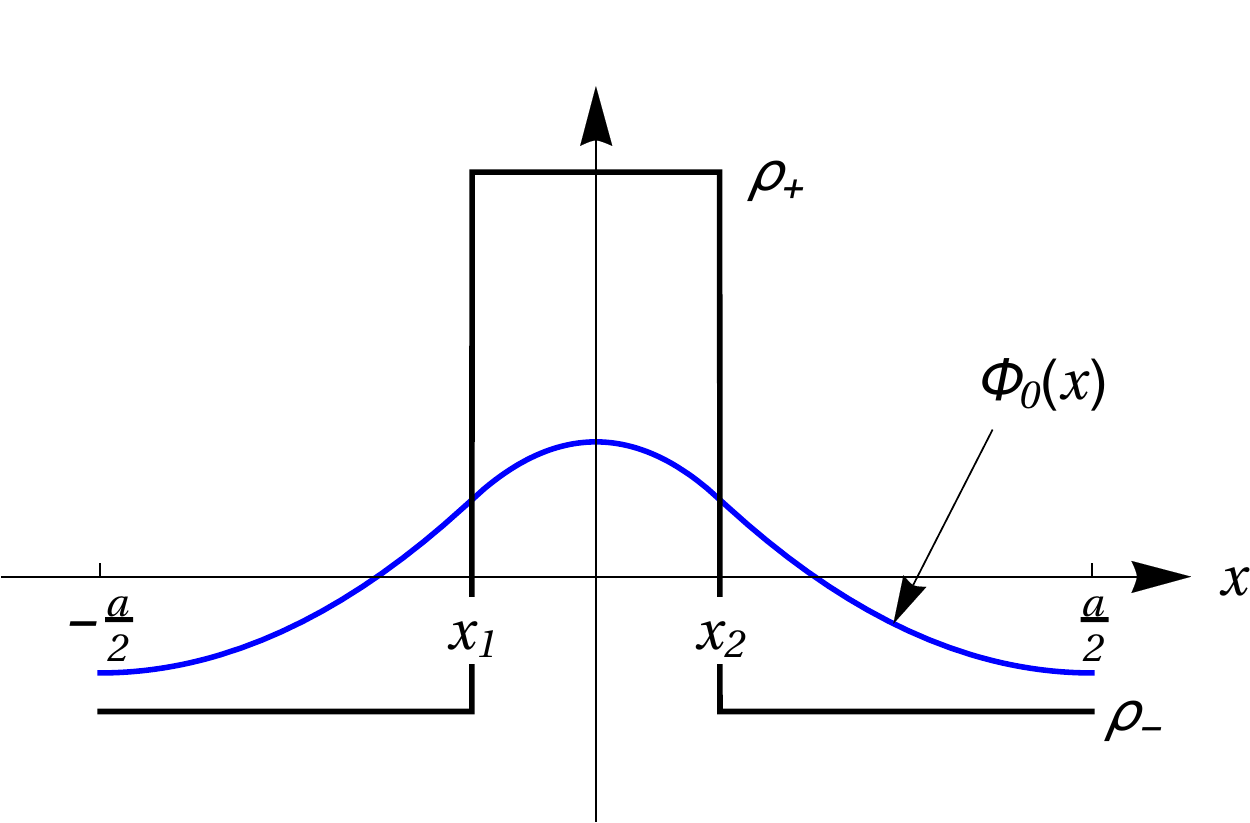}
\caption{The unperturbed charge distribution and its  potential $\Phi_0$.}
\label{fig-lasagna}
\end{figure}
Let us consider  a proton cluster $\cP$ in the shape of one slab placed in the center of unit cell with size $a,b,c$ and
volume $V_\cC\! =\! abc$.
The slab is perpendicular to the $x$-axis and occupies a $w$ fraction of the total cell volume, $0<w<1$. The charge density 
contrast between phases is $\Delta \rho = \rho_+ - \rho_-$, so the slab is positively charged with the density
 $\rho_+ = (1-w) \,\Delta \rho$ and immersed in negatively charged neutron 
gas $\rho_-= -w \, \Delta \rho$ (we assume the homogeneous electron background).
For such charge distribution, the unperturbed potential $\Phi_0$, Fig.~\ref{fig-lasagna}, is a function  of $x$ only and takes the form
\begin{align}
&\Phi_0(\vb{r}) = \nonumber \\
 = & \; \pi\Delta \rho
\begin{cases}
     w \left( \frac{1}{6} a^2 \left(w^2+2\right)+ 2 a x+ 2 x^2\right)
     & -\frac{a}{2}\leq  x< x_1 \\
  (w-1) \left( \frac{1}{6} a^2 (w-2) w + 2 x^2\right) 
     & ~ x_1 \leq  x< x_2  \\
    w \left( \frac{1}{6}  a^2 \left(w^2+2\right)- 2 a x+ 2 x^2\right)
     & ~ x_2 \leq   x\leq \frac{a}{2} ~,
\end{cases}  
\label{phi0}
\end{align}
where $x_{1,2} = \pm\frac{aw}{2}$ correspond to the positions of  slab edges.
Such configuration fulfill the necessary conditions for minimum, i.e. equations coming from vanishing variation 
of total energy in the first order. The sufficient condition for minimum is expressed by the 
positive value of total energy variation  in the second order $\dd{2}\tilde{\ep}$ where the constraints 
for baryon number and charge conservation are imposed. Such energy  variation
for any deformation of the proton cluster $\eps$ is expressed by the integral over
proton cluster surface~$\dP$~\cite{Kubis:2016fmw} 
\beq
\begin{array}{ll}
\dd{2}\tilde{\ep} =  & \\ 
\fracd{1}{2 V_\cC} \int_\dP \left(\sigma ((\nabla \eps)^2 - B^2 \eps^2) + 
\Delta\rho \;(\pa_n\Phi_0\; \eps^2 +\delta_\eps\Phi \;\eps) \right)  dS , &  
\end{array}
\label{envar2} 
\eeq
where $B^2 = \kappa_1^2 + \kappa_2^2 = 0$ is the sum of squared principal curvatures, $\pa_n\Phi_0$  is the normal derivative
of unperturbed potential, $\delta_\eps\Phi$ is the first order perturbation of the potential caused by the surface deformation
$\eps$. One must remember that here we consider only the  deformation which preserves the volume of the cluster, which means
\beq
\int_\dP \eps \; dS = 0 .
\label{vol-preserv}
\eeq
 Deformations of this kind 
may be expressed in terms of Fourier series on the $y,z$-plane.
The deformations on each slab face are independent,  
so we get two series for each face located at $x_1$ and $x_2$. 
For further calculations it is convenient to introduce an
expansion based on complex amplitudes  $\alpha^j_{mk}$, where $m$ and $k$ are the mode indices and the superscript $j$
 corresponds to  the face number
\beq
\eps^j(y,z) = \sumprime_{m,k=-\infty}^\infty \alpha_{mk}^j \exp(\ii \vb{K}_{0mk}\cdot \vb{x} )~,
\label{expansion}
\eeq
where we introduce the 3-dimensional discrete wave vector 
\beq
\vb{K}_{nmk}=\left(\frac{2\pi n}{a},\frac{2\pi m}{b},\frac{2\pi k}{c}\right)
\eeq
and $\vb{x} = (x,y,z)$. The $m$ and $k$ indices 
take integer values from $-\infty $ to $+\infty $ except the case when both of them equal zero, which is indicated 
by an apostrophe in the sum sign. The lack of $(0,0)$-mode is consistent with the volume 
conservation condition, Eq.~(\ref{vol-preserv}). The inclusion of such compression modes would require taking into account the particle density perturbation. Their stability is controlled mainly by the volume compressibility coefficient
 $K_V=\der{P}{\rho}$ -  quantity being dependent on the details of nuclear interactions.
The compression modes may be carried out separately as the $(0,0)$-mode is orthogonal to the shape changing modes. 
As we are interested only in the shape stability, we postpone the compression modes for future work. 
 
\begin{figure}[h!]
\centering
\includegraphics[width=.8\columnwidth]{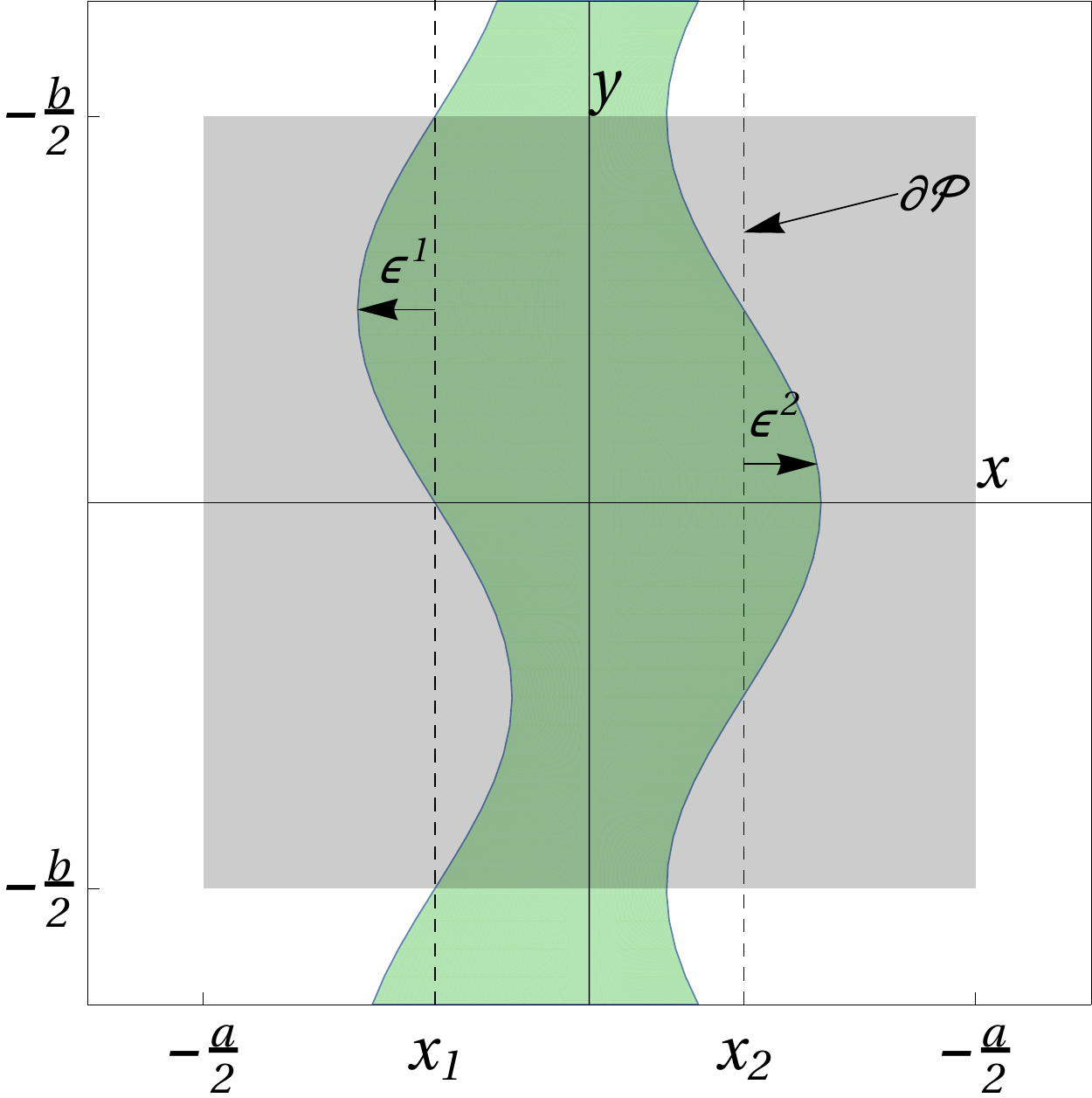}
\caption{The slab deformation in the unit cell (the grey rectangle). The first face is only with sine mode whereas the second 
face with  cosine mode.}
 \label{fig-mode}
\end{figure}

Since the deformation function $\eps^j(y,z) $ must be real it means that the complex amplitudes should fulfill
the following relations
\begin{equation}
(\alpha_{mk}^j)^* = \alpha_{-m,-k}^j~ .
\label{conjugation}
\end{equation} 
It is more natural to introduce the  cosine $\eps_{mk,C} \cos(\frac{m y}{b}+\frac{k z}{c})$ and sine modes 
$ \eps_{mk,S} \sin(\frac{m y}{b}+\frac{k z}{c}) $ in the expansion  Eq.~\ref{expansion}. Then the relation between complex and real amplitudes is
\begin{align}
   \eps_{mk,C}^j & =  \alpha_{mk}^j+\alpha_{-m,-k}^j \\
\eps_{mk,S}^j  & =  \ii(\alpha_{mk}^j  - \alpha_{-m,-k}^j)~ .
\end{align}
In the Fig.~\ref{fig-mode} an example of the slab deformation with real amplitudes taking the values
$\eps_{10,S}^1 = \eps_{10,C}^2 \neq 0$ and  $\eps_{10,C}^1 = \eps_{10,S}^2 =0$ is shown.
Below we present and discuss the subsequent contribution to the total energy variation $\dd{2}\tilde{\ep}$, 
Eq.~(\ref{envar2}).
In terms of complex amplitudes the surface energy contribution to the 2nd order variation takes simple form
\beqa
\dd{2}\tilde{\ep}_S &=& \frac{\sigma }{2 V_\cC} \int_\dP  \left(  (\nabla \eps)^2 - B^2 \eps^2 \right)  \; dS \nonumber \\ 
& = &  \frac{\sigma}{2 a}\sum_{j=1}^2 \sumprime_{mk} \alpha_{mk}^j \alpha_{-m,-k}^j K_{0mk}^2~. \label{var-surf}
\eeqa
It is always positive, as $B^2=0$ for the slab and due to relations (\ref{conjugation})
we get $\alpha_{mk}^j \alpha_{-m,-k}^j = |\alpha_{mk}^j|^2>0$. That is an important fact, because  in many cases the surface energy
acts as a destabilizer, like, for example, $\dd{2}\tilde{\ep}_S <0 $ in the case of  Rayleigh-Plateau instability.

The contribution coming from the electrostatic interaction includes two terms. The first one, determined by the normal derivative of potential $\Phi_0$, is given by
\beqa
\dd{2}\tilde{\ep}_{norm}&=& \frac{\Delta\rho}{2 V_\cC} \int_\dP   \; \pa_n\Phi_0\; \eps^2  \; dS \nonumber \\ 
& =& \frac{\Delta \rho}{2 a} \sum_{j=1}^2  \sumprime_{mk} \pa_n \Phi_0(x_j)\; \alpha_{mk}^j \alpha_{-m,-k}^j ~.
\label{var-norm}
\eeqa
The normal derivatives at the slab faces are $ \pa_n\Phi_0(x_1)=\pa_n\Phi_0(x_1)=-2\pi (1-w)w a \Delta\rho $.
So, finally one gets
\begin{equation}
\dd{2}\tilde{\ep}_{norm} = -\pi \Delta\rho^2 \,w(1-w) \sum_{j=1}^2  \sumprime_{mk} \alpha_{mk}^j \alpha_{-m,-k}^j  
\end{equation}
which is always negative.
The second term of the Coulomb interaction part is associated with perturbation of the potential $\delta_\eps\Phi$ 
\beq
\dd{2}\tilde{\ep}_\Phi = \frac{\Delta\rho}{2 V_\cC} \int_\dP \delta_\eps\Phi \;\eps \; dS ,
\eeq
where $\delta_\eps\Phi$ is the first order potential  perturbation calculated thanks to the periodic Green function
\beq
\delta_\eps\Phi(\vb{x}) = \Delta\rho\int_{\dP} G_P(\vb{x},\vb{x}') \eps(\vb{x}') \; dS' ~.  
\label{var-potential}
\eeq
For the  three-dimensional unit cell with sizes $a,b,c$ the periodic Green function can be expressed as the sum over discrete modes numbered by three indices $n,m,k$ 
\beq
G_{P}(\vbx,\vbx') = \frac{4\pi}{abc} \sumprime_{n,m,k=-\infty}^{+\infty}
\frac{\exp({\ii \vb{K}_{nmk}\cdot(\vbx-\vbx')})}{K_{nmk}^2} ~.
\label{green}
\eeq    
The prime sign in the summation means that $n,m,k$ cannot vanish simultaneously, similarly as in the Eq.~(\ref{expansion}). Such Green function fulfills the Poisson equation in the unit cell with periodic boundary conditions \cite{marshall,tyagi}
(we use CGS units)
\beq
\nabla^2 G_P(\vbx,\vbx') = - 4\pi\left(\delta(\vbx-\vbx')-\frac{1}{abc}\right)
\eeq
and the absence of (0,0,0) mode in the expansion (\ref{green}) is a consequence of the unit cell neutrality. By joining Eqs.(\ref{expansion},\ref{var-potential},\ref{green}) we obtain the change in the energy
with respect to the potential variation
\beq
\dd{2}\tilde{\ep}_\Phi = 2\pi\Delta\rho^2 \sum_{i,j=1}^2  \sumprime_{m,k} \alpha^i_{mk}\,\alpha^j_{-m,-k}
F(\xi_{ij}\, ,\chi_{mk}) ~,
\label{var-pot}
\eeq
where $\xi_{ij}$ corresponds to the difference between the faces location
\beq
\xi_{ij} = \frac{x_i-x_j}{a}
\eeq
and $\chi_{mk}$ is the norm of the  dimensionless wave vector $\bm{\chi}_{mk} = (\frac{2\pi m a}{b} , \frac{2\pi k a}{c} )$ 
\beq
\chi_{mk} = a K_{0mk} =\sqrt{ \left(\frac{2\pi m a}{b}\right)^2 + \left(\frac{2\pi k a}{c}\right)^2}~.
\eeq
The vector $\bm{\chi}_{mk}$ describes the manner in which the  face is vibrating - the indices $m$ and $k$ determine the number of wavelengths being placed in the face sizes $b$ and $c$.
The function $F(\xi,\chi)$  comes  from the summation over the $n$ index. 
As the $n$-numbered modes do not depend on $x$, the summation over $n$ can be done separately and defines
a function $F$  
\beq
F(\xi,\chi) \stackrel{\mathrm{df}}{=}  \sum_{n=-\infty}^{\infty} \frac{\mathrm{e}^{\ii 2\pi n \xi}}{(2\pi n)^2 +\chi^2} ~.
\eeq
The above series can be evaluated \cite{gradshteyn} to a closed form
\beq
F(\xi,\chi) = \frac{ \cosh \left( \frac{\chi}{2} 
   (1-2\left| \xi \right| )\right)}{2 \chi \sinh(\frac{\chi }{2}) }~.
\eeq
The function is positive for any $\xi$ and $\chi$. In some special cases the function 
simplifies to
\begin{align}
F(0,\chi) &= \frac{\coth \left(\frac{\chi }{2}\right)}{2 \chi }~, \\
F(\oht,\chi)& = \frac{1}{2 \chi  \sinh(\frac{\chi}{2})} ~.
\end{align}
 Taking together all energy terms we get the total energy variation
 expressed in terms of mode amplitudes $\alpha_{mk}$
\begin{align}
\dd{2}\tilde{\ep} = \sum_{i,j=1}^{2} \sumprime_{m,k} &
\left\{ \frac{\sigma}{2 a^3}\, \chi^2_{mk}\delta_{ij}  - \pi \Delta\rho^2 w (1-w) \delta_{ij}\; + \right. \\ \nonumber
  & \left.  ~~~ + \;2  \pi\,\Delta\rho^2 F(\xi_{ij}\, ,\chi_{mk})\, \right\}  \alpha^i_{mk}\alpha^j_{-m,-k}~.
\end{align}
As the $\dd{2}\tilde{\ep}$ is the 
quadratic form of the amplitudes $\alpha_{mk}$ the competition between the terms in curly brackets 
decides about the stability of the face surface. 
There are three characteristic terms: one from the surface energy being positive and two from
Coulomb interactions: the first is negative and the second is always positive. 
It seems that  the stability consideration depends on the values of surface tension $\sigma$ or charge contrast $\Delta\rho$
but it appears that these parameters may be removed from our analysis. One of the conditions for the minimum of the energy 
for unit cell is the virial theorem   \cite{Kubis:2016fmw}. The theorem takes the form of relation between the surface and Coulomb energy of the cell. For the cell with high symmetry, it has the simple form
\beq
E_S = 2 E_{Coul} ~,
\eeq
from which we may get the relation between $\sigma$ and $\Delta\rho$
\beq
\sigma =\frac{1}{6} \pi  a^3 \Delta \rho ^2 (w-1)^2 w^2~.
\eeq
Finally the total energy variation  may be written as the quadratic form of $\alpha^i_{km}$
\beq
\dd{2}\tilde{\ep} = \pi \Delta\rho^2 \sum_{i,j=1}^{2} \sumprime_{m,k} \Ac^{ij}_{mk}  \alpha^i_{mk}\alpha^j_{-m,-k}
\label{final-ep}
\eeq
with its coefficients $\Ac_{mk}^{ij}$ given by 
\begin{align}
\Ac_{mk}^{ij} =  & \left\{ (\fract{1}{12} w^2 (1-w)^2 \chi _{mk}^2 - w (1-w))\,
  \delta _{ij} \right. \nonumber \\ 
 & \left. + \;2\, F(\xi _{ij},\chi _{mk}) \right\} .
 \label{A-mk}
\end{align}
The obtained general  expression, Eq.~(\ref{final-ep}) for the second order variation of the total energy
allows us to determine whether the slab is stable with respect to any  deformations
preserving its volume. 

It is worth noting that $\Ac_{mk}^{ij}$ coefficients, which decide about stability, do not depend on 
the strength of interactions being determined by surface tension  $\sigma$ and charge contrast $\Delta\rho$.
In this way, we obtained an interesting result that the stability of pasta depends only on the geometry of phase and mode
under consideration and not on the details of strong or electromagnetic interactions.

Before  general discussion of the stability of a single slab we show how the above results work in 
the  stability analysis for particular class modes.

\section{An example of stability analysis}

\label{sect-example}

\begin{figure}[t]
\centering
\includegraphics[width=.5\textwidth]{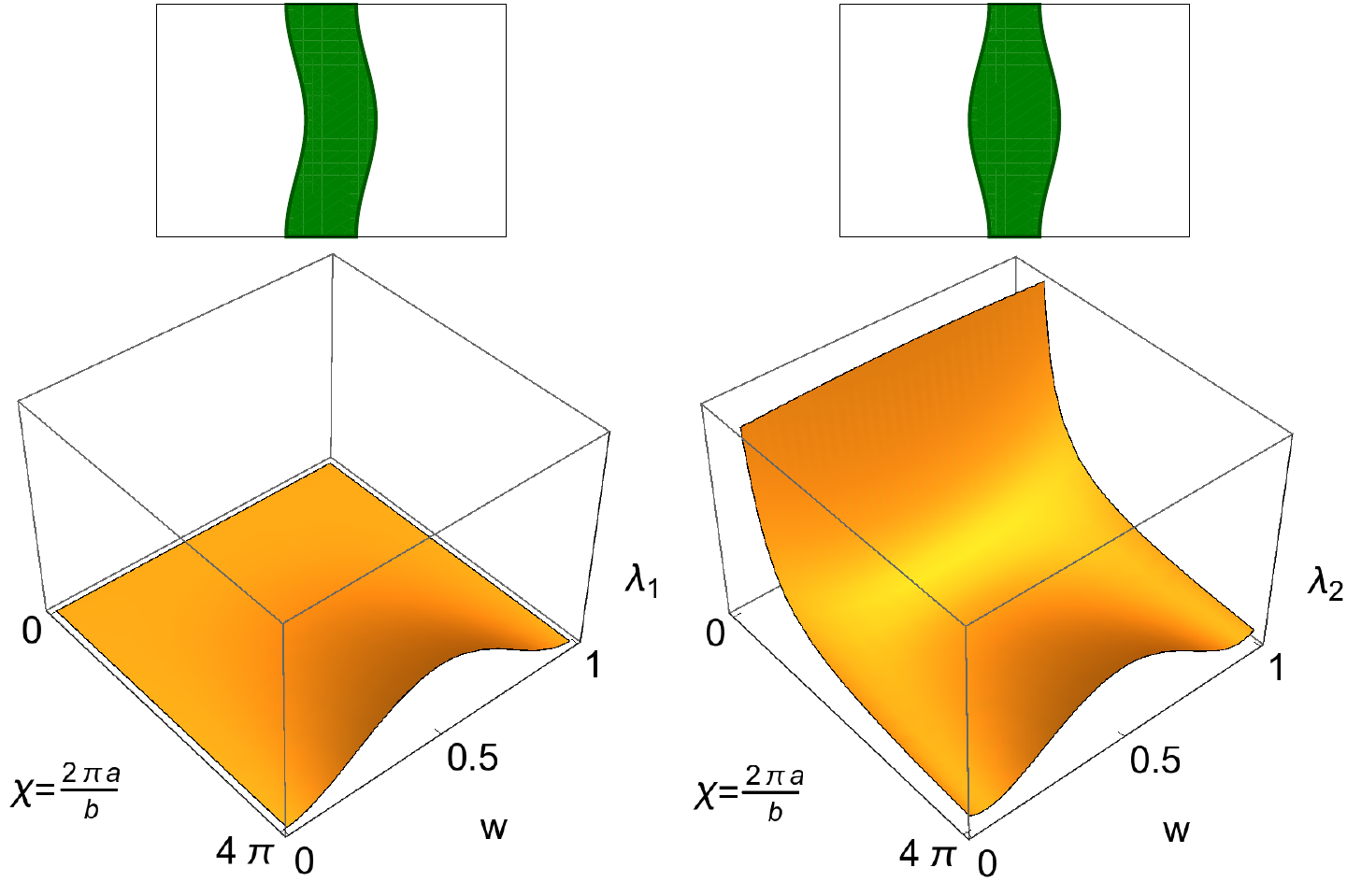}
\caption{The  stability functions $\lambda_i(w,\chi)$ for two types of deformations: snaky (left) and
 hourglass-shaped (right) mode.}
\label{plotstab2}
\end{figure}
Let us  consider the simplest surface perturbation consisting of combination of 
sine and cosine modes going along the $y$-axis for each face, which corresponds to $m\!=\! \pm 1$ and $k\!=\!0$.
Then, for the {\it i}-th face, the only non-vanishing complex amplitudes $\alpha^i_{mk}$ are:
\begin{align}
\alpha^i_{~10} &= \frac{1}{2} \left(\epsilon^i_{C}-\ii\, \epsilon^i_{S}\right)~,\\
\alpha^i_{-10} &= \frac{1}{2} \left(\epsilon^i_{C}+\ii\, \epsilon^i_{S}\right)~,
\end{align}
where we introduced the real amplitudes $\eps^i_S$ and $\eps^i_C$ corresponding to the functions 
$\sin(\frac{2\pi y}{b})$ and $\cos(\frac{2\pi y}{b})$. The deformation $\eps^i$ for the {\it i}-th
face is then a function of $y$ only
\beq
\eps^i(y) = \epsilon^i_{C}\cos(\frac{2\pi y}{b}) + \epsilon^i_{S}\sin(\frac{2\pi y}{b}) ~.
\eeq 
It is convenient to introduce the vector $\vb{\epsilon}$ built of deformation amplitudes 
\beq
\vb{\epsilon} =(\epsilon_C^1,\epsilon_C^2,\epsilon_S^1,\epsilon_S^2)~.
\eeq
Then the total energy variation for such deformation may be written down in the matrix form
\beq
\dd{2}\tilde{\ep}= \pi \drho^2\vb{\eps} \, \vb{\hat{M}}\, \vb{\eps}^{\, T}~,
\eeq
where the dimensionless matrix $\vb{\hat{M}}$ is
\beq
\vb{\hat{M}} = \left(
\begin{matrix}
 A & B & 0 & 0 \\
B & A & 0 & 0 \\
 0 & 0 & A & B \\
 0 & 0 & B & A \\
\end{matrix}
\right)
\label{M-matrix}
\eeq
and its elements are:
\begin{align}
A&=\fract{1}{24}(1-w)^2 w^2 \chi^2  - \oht w(1-w) +F(0,\chi) ~, \\ \nonumber
B&= F(w,\chi)~,
\label{ABeqn}
\end{align}
where $w$ is the volume fraction and $\chi$ determines the wavelength of the mode in comparison to the cell size $\chi = 2\pi a/b$.
The inspection of eigenvalues and eigenvectors of $\vb{\hat{M}}$ allows for a complete stability analysis for the 
deformation we have chosen.
The matrix $\vb{\hat{M}}$ given by Eq.~(\ref{M-matrix}) possesses the two-fold degenerated two eigenvalues.
Further, we call the  $\vb{\hat{M}}$-eigenvalues $\lambda_l~,~l=1\dots 4$ as the stability function.
For our concrete form  
of $\vb{\hat{M}}$ the eigenvalues and their eigenvectors are
\begin{align}
\lambda_{1,3} & = \fract{1}{24}(1-w)^2 w^2 \chi^2  - \oht w(1-w) + F(0,\chi) - F(w,\chi) \nonumber \\[.5em]
\vb{\eps}_{1} & = (-1,1,0,0)~~,~~ \vb{\eps}_{3} = (0,0,-1,1) \nonumber \\[.8em]
\lambda_{2,4} & = \fract{1}{24} (1-w)^2 w^2 \chi^2  - \oht w(1-w) + F(0,\chi) + F(w,\chi) \nonumber \\[.5em]
\vb{\eps}_{2} & = (1,1,0,0)~~,~~ \vb{\eps}_{4} = (0,0,1,1)~.\nonumber \\ 
\end{align}
The stability functions $\lambda_i$ do not depend on the details of interactions $\sigma, \drho$ but only on the geometry
of our system which is described by volume fraction $w$ and mode wavelength $\chi$.
  In Fig.~\ref{plotstab2} the stability functions are plotted in the $w,\chi$ parameter space for their eigenvectors.
These vectors represent  two classes
of modes. We may call them as snaky and hourglass-shaped modes. The snaky modes occur when the deformations on 
both faces are in phase ($\vb{\eps}_{1},~\vb{\eps}_{3}$)  and hourglass modes occur when the deformations on faces are out of phase  ($\vb{\eps}_{2},~\vb{\eps}_{4}$).
As we may notice, for all values of volume fraction  and mode wavelengths the stability functions
 are positive, which means, that for both classes of modes the slab is stable. 
However the stability is not too strong.
Careful inspection of $\lambda_1$ and $\lambda_2$ shows that those functions go to~0 when $\chi$ and $w$ 
approach to
some specific values
\beq
\lambda_1|_{\chi\rightarrow 0} \rightarrow 0 {\rm ~~for ~any~} w
\label{asympt1}
\eeq
and
\beq
\lambda_2|_{\chi\rightarrow \infty} \rightarrow 0 {\rm~ for~} w\rightarrow 0 {\rm ~or~}1~.
\label{asympt2}
\eeq
It means that snaky modes become unstable in the limit of very long waves regardless of slab thickness, whereas the hourglass modes become unstable for very thin slab and very short waves.  To sum up, we may say the slab becomes asymptotically unstable
for very long mode or for very short mode when the slab becomes very thin in comparison to unit cell width.
{One should note that, the case when $w \approx 0~ {\rm or} ~w \approx 1$ must be treated with caution 
because in reality the cluster surface has finite thickness and for very thin slab the validity of the liquid drop model could be questioned.}
 Nevertheless, the first case of asymptotic instability , Eq.(\ref{asympt1}) , is worthy of careful inspection as it is connected to the macroscopic deformation and allows for determination of elastic properties of lasagna phase, which is shown in the next section.

\section{Elastic properties}

Nuclear pastas share their elastic properties with liquid crystals.
If the wavelength of the  snaky mode is very large in comparison to the cell size, it corresponds to the so-called splay deformation of the liquid crystal.  
It allows to determine elastic constant $K_1$ for the lasagna phase what was shown by Pethick and Potekhin in \cite{Pethick:1998qv}. 
By definition, the constant $K_1$, relates the deformation energy with the transverse derivative of the deformation field
when the mode wavelength becomes very large. In our notation
the relation takes the form 
\[
\delta^2\tilde{\ep}=\frac{K_1}{2}\langle (\pa_y^2\eps)^2 \rangle~,
\]
where the brackets $\langle ..\rangle$ mean the average over the slab surface.
Taking the definition of $K_1$ in the limit of the very long mode, we get 
\beq
K_1 = 4 a^4\lim_{\chi\rightarrow 0}\frac{\delta^2\tilde{\ep}}{\chi^4 \eps^2}~,
\label{K1}
\eeq
(here $\eps$ denote the mode amplitude only). 
The deformation energy for the snaky mode is 
\begin{align}
\delta\tilde{\ep}_{snaky}\; = \; & \pi \Delta\rho^2  \left( 
   \frac{1}{12} (1 - w)^2 w^2 \chi^2 - (1 - w) w \right.  \nonumber \\
  &  \left.  +\; 2\; (F(0,\chi) - F(w,\chi)) \frac{}{}\right) \eps^2 ~.
\end{align}
Applying the limit in Eq.(\ref{K1}) we get 
\beq
K_1= \frac{1}{180} \pi  a^4 \Delta \rho ^2 (w-1)^2 w^2 \left(1+2 w -2 w^2\right)~,
\eeq
which exactly corresponds to the  result   of \cite{Pethick:1998qv} if the $\Delta\rho$ is 
replaced by the unperturbed Coulomb energy of the cell $\ep_{C,0} =\frac{1}{6} \pi  a^2 \Delta \rho^2 (1-w)^2 w^2$.
In comparison to  \cite{Pethick:1998qv} our approach represents an improvement.
\corr{
Pethick and Potekhin used a triple sum for the deformation energy, we mean Eq.(8) in \cite{Pethick:1998qv}, which gives correct result only in the limit of the very long mode $\chi\rightarrow 0$. 
In fact, that series represents an asymptotic expansion in the powers of mode wavelengths 
and the leading term gives correct result. However the expansion is divergent for finite  $\chi$.
The detailed discussion of this divergence is shown in the Appendix.
The approach, presented here,  allows to avoid any divergences and is not limited to the case $\chi\rightarrow 0$.}

\section{General discussion of stability}

In the Section \ref{sect-example}  the stability analysis for the simplest slab deformation was carried out. 
The full stability analysis would require the inclusion of modes for all multiplicities $m$ and $k$.
Writing down the matrix $\vb{\hat{M}}$ for all modes it appears to take the block-diagonal form 
\beq
\vb{\hat{M}} = \left(
\begin{matrix}
\ddots & & & & & \\
&A & B & 0 & 0 &\\
&B  & A & 0 & 0 &\\
&0  & 0 & A & B &\\
&0 & 0 &  B & A & \\
& & & & & \ddots \\
\end{matrix}~
\right) ~,
\eeq
where in the $mk$-th position we get the same matrix as in Eq.~(\ref{M-matrix}) with $A$ and $B$ elements  taking
 the same form as  in Eq.~(\ref{ABeqn}) but with the replacement
$\chi \rightarrow \chi_{km}$. Such a form of $\vb{\hat{M}}$ means that the modes with different multiplicity $m,k$ and
the cosine-like and sine-like modes do not couple. So, taking fixed $m,k$ one may repeat the discussion
from the Section \ref{sect-example}
and finally  conclude that the single slab is stable for any mode keeping in mind the asymptotic cases, 
described by the  Eqs.~(\ref{asympt1},\ref{asympt2}). 
\label{sec-general}

\section{Multi-slab modes}

\begin{figure*}
\centering
\includegraphics[width=.99\textwidth]{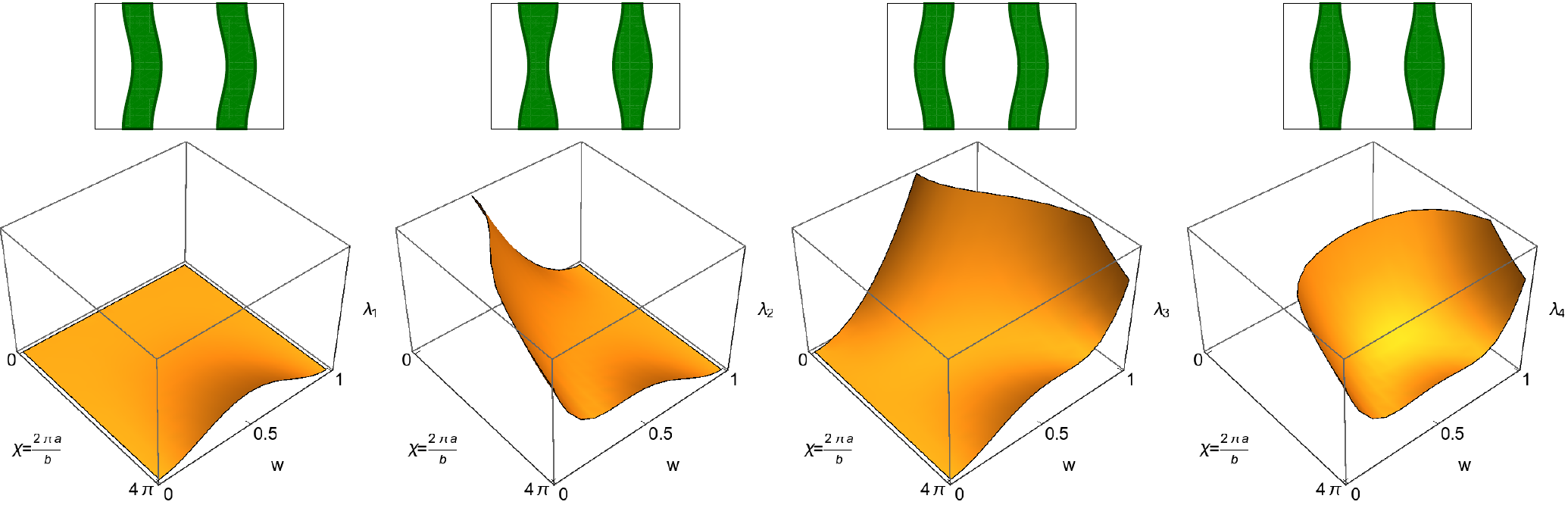}
\caption{The  slabs deformations and their stability functions $\lambda_i(w,\chi)$ in the case of two slabs per cell.}
\label{plotstab4}
\end{figure*}

In previous sections the cell with only one proton layer was considered. 
However, one may consider many slabs which are placed  in a cell with periodic 
boundary conditions. Such system allows to test a larger class of perturbations and makes
our analysis of stability more complete. Let us suppose we have $N$ slabs in one cell.
The expressions derived earlier for energy variation (\ref{var-surf},\ref{var-norm},\ref{var-pot})
take the same form, except the summation over the surfaces which now takes the range 
 $i,j=1..2N$, as we have $2N$ surfaces for $N$ slabs. Moreover, the value of the normal derivative of the electric potential
 scales with the number of slabs $N$ according to the rule 
 \beq
  \pa_n\Phi_0(x_i)=\frac{-2\pi}{N} (1-w)w a \Delta\rho .
  \eeq
As was discussed in section \ref{sec-general} it is enough to test the only one type of mode per surface.
 Then the stability matrix $\vb{\hat{M}}$ for one given mode has dimension $2N\times 2N$.
As an example we show results for the case of two slabs, $N=2$ and cosine-mode with $m=1,k=0$.
 Then we have four  distinct eigenvalues of  stability matrix  $\vb{\hat{M}}$  :
\begin{equation}
\begin{array}{c}
 \lambda_1 = D+F\left(\frac{1}{2},\chi \right)-F\left(\frac{w}{2},\chi
   \right)-F\left(\frac{w+1}{2},\chi \right) \\
 \lambda_2 = D-F\left(\frac{1}{2},\chi \right)+F\left(\frac{w}{2},\chi
   \right)-F\left(\frac{w+1}{2},\chi \right) \\
  \lambda_3 = D-F\left(\frac{1}{2},\chi \right)-F\left(\frac{w}{2},\chi
   \right)+F\left(\frac{w+1}{2},\chi \right) \\
  \lambda_4 = D+F\left(\frac{1}{2},\chi \right)+F\left(\frac{w}{2},\chi
   \right)+F\left(\frac{w+1}{2},\chi \right) \\
\end{array}
\end{equation}
where  $D$  is
\begin{equation}
D = \frac{1}{192} (w-1)^2 w^2 \chi ^2+\frac{1}{4} (w-1) w + F(0,\chi ) ~.
\end{equation}
All of these eigenvalues are positive. Their dependence on volume fraction $w$ and mode wavelength $\chi$
and the corresponding slab deformations are shown in the Fig.\ref{plotstab4}.
As one may see, the eigenmodes of the deformation  of multi-slab system
are always the combination of snaky and hourglass modes.

\section{Conclusions}
In this work, by use of analytical methods, we have shown that proton  clusters having the form of slabs placed  periodically in space are stable 
for all values of volume fraction occupied by the cluster. It is a compelling result. All works based on CLDM (see for example \cite{ravenhall83,Oyamatsu:1993zz,Williams:1985prf,Nakazato:2009ed}),
show that  different shapes of  pasta   are preferred for different values of $w$.  
Here, we have shown that the lasagna phase is stable in the whole range of $w$.
One must remember that our analysis means that  the lasagna phase represents merely a local minimum. The transition to another 
geometry is not totally  blocked, but requires finite size deformation in order to exceed the energy barrier.
It may be interpreted as the fact that, at least for some range of volume fraction the lasagna phase is metastable.
That could be quite interesting for the dynamics
of pasta appearance during the neutron star formation.

Our analysis is based on small deformations so it cannot state at which range of $w$ the lasagna represents global minimum of the cell energy.
First, the global minimum statement requires the knowledge of exact solutions of Eq.~(\ref{H-eqn}) 
for other kind of  shape than flat slab. So far, we have not known such 
solutions  in the CLDM approach. The seeking of them marks out the direction of further research of pasta by  
differential geometry methods.

We are also conscious that the approach, based on the CLDM, has its limitations and the inclusion of such effects
like finite thickness of the cluster surface  or the temperature fluctuations could change the final 
conclusion concerning lasagna phase stability.


\corr{

\section*{Appendix}

\setcounter{equation}{0}
\renewcommand\theequation{A.\arabic{equation}}

The Coulomb energy in the work \cite{Pethick:1998qv}, Eq.(8),  was finally expressed by the power  series
\begin{equation}
{\cal S}=\sum_{j=1}^\infty (-1)^j \;a_j(\xi)\;(k r_c/\pi)^{2j} ~ ,
\label{pethick-series}
\end{equation}
here we keep the notation used by Pethick and Potekhin: $2 r_c$  is the cell size, $k$ is the mode wavenumber and $\xi$ is the dimensionless  amplitude of perturbation.  Coefficients $a_j(\xi)$ are determined by the double sum
\begin{equation}
a_j(\xi)= \frac{8}{\pi^3 w^2}\sum_{m=1}^\infty \frac{\sin^2(m \pi w)}{m^{4+2j}} \sum_{n=1}^\infty n^{2j} J_n^2(m \xi)~.  
\label{coeff}
\end{equation}
The series given by $\cal S$ is divergent.
It is enough to take only one term from the double sum, Eq.(\ref{coeff}) for $m=1$ and $n=j/2$ and then every coefficient  $a_j(\xi) $ is bounded from below by the expression
\begin{equation}
\frac{8}{\pi^3 w^2}\left(\frac{j}{2}\right)^{2 j}  
     \sin ^2(\pi  w) J_{\frac{j}{2}}^2(\xi) ~ <  ~ a_j(\xi) ~.
  \end{equation}
  The $\frac{j}{2}$-th Bessel function for small amplitude ($\xi<1$),  again, may be bounded by its Taylor expansion  
\begin{equation}
\frac{ (\xi/2)^\frac{j}{2} (1+\frac{j}{2}-(\xi/2) ^2)}{(\frac{j}{2}+1)!} ~ <  ~ J_\frac{j}{2}(\xi)  ~.
\end{equation}
So, finally, all the coefficients of the power series $\cal S$ are underestimated by the new ones $\tilde{a}_j$ 
\[
\tilde{a}_j(\xi) = \left(\frac{e \xi j}{4}\right)^j 
\frac{(-2 j+\xi ^2-4)^2 \sin ^2(\pi  w)}{4 \pi j(j+2)^2} ~ < ~ a_j(\xi)~,
\]
where we removed factorial by use of the  Stirling formula. 
The power series with coefficients $\tilde{a}_j$ is however divergent  because for large $j$ the $\tilde{a}_j$
 behaves like $j^j$ and the convergence radius is zero
 \begin{equation}
 \displaystyle  r_{conv}=\lim_{j\rightarrow\infty}\;\frac{\tilde{a}_j}{\tilde{a}_{j+1}} = 0~.
 \end{equation}
This divergence comes from the introduction of the sum over $j$ in the expression for
 Coulomb energy. The Coulomb energy, was initially  expressed by the convergent double sum over $m,n$ indices, Eq.(7) of \cite{Pethick:1998qv}, then the third summation over  $j$ was introduced 
in the following way (screening was neglected $k_{TF}=0$)
\begin{equation}
\frac{1}{(m\pi/r_c)^2+(n k)^2} = 
 \left(\frac{r_c}{m\pi}\right)^2\sum_{j=0}^\infty (-1)^j\left(\frac{n k r_c }{m \pi}\right)^{2j}
 \label{geom-exp}
  \end{equation}
 and then the order of summation sequence was changed
$\sum_{n,m}\sum_j \rightarrow \sum_j\sum_{n,m} $.
 The change in the summation order is allowed only if all sub-series are convergent, however, the expansion given by Eq.(\ref{geom-exp})  is convergent only if
\begin{equation}
n< \frac{\pi}{k r_c} m~ .  
\end{equation}
That means that the indices are not independent: the summation sequence is not interchangeable. The authors of \cite{Pethick:1998qv} got a correct result because they took the double limit:
$\xi\rightarrow 0$ (small amplitude of deformation) and 
$ \frac{\pi}{k r_c} \rightarrow \infty$ (very long mode).
Summarizing, we want to emphasize that  such expression for the Coulomb energy, Eq.(8) in \cite{Pethick:1998qv}, is valid  only in the limit of the very long mode, 
$k r_c \rightarrow  0$ .

}

\end{document}